\documentclass[%
 reprint,
 amsmath,amssymb,
 aps,
 prl,
 longbibliography,
 lengthcheck,%
]{revtex4-1}

\usepackage{graphicx}
\usepackage{dcolumn}
\usepackage{bm}
\usepackage{hyperref}

\begin{document}

\preprint{APS/123-QED}

\title{Classical microscopic derivation of the relativistic hydrodynamics equations}

\author{P. A. Andreev}
\email{andreevpa@physics.msu.ru}
 \affiliation{Department of General Physics, Physics Faculty, Moscow State
University, Moscow, Russian Federation.}

\date{\today}

\begin{abstract}

We present microscopic derivation of the relativistic
hydrodynamics (RHD) equations directly from mechanics omitting
derivation of kinetic equation. We derive continuity equation and
energy-momentum conservation law. We also derive equation of
evolution of particles current. In non-relativistic hydrodynamics
equation of particles current evolution coincide with the equation
of momentum evolution, the Maxwell's equations contain
concentration and electric current (which proportional to the
particles current), so, to get a close set of equations we should
have equations of evolution of the concentration and the particles current.
Evolution of the particles current depends on the electrical and magnetic
fields. Thus, we obtain the set of the RHD equations as the set of the
continuity equation, the equation of particles current and the Maxwell
equations. This approximation does not require to include
the evolution of momentum and allows to consider physical problems. Certanly, particles current evolution equation contains some new
functions which we can express via concentration and particles
current or we can derive equation for this functions, and, thus,
get to more general approximation. This approximation also
developed and discussed in this paper.
\end{abstract}

\maketitle


\section{\label{sec:level1} I. Introduction}

Relativistic kinetics and hydrodynamics have been in a center of
attention. A lot of papers and books dedicated to this topic have
been published in last several years. Wide recent review of relativistic
kinetics has been presented in R. Hakim's book ~\cite{Hakim book
11}, but we also should mention a paper about relativistic
kinetics ~\cite{Kuz'menkov 91}, where was suggested a method of
equations derivation which we suppose to use for derivation of
hydrodynamics equations.

When one considers the relativistic hydrodynamics equations he usually
uses conservation laws: conservation of particles number and
momentum and energy, in Lorentz covariant form, along with the
Maxwell equations. To get a close set of the hydrodynamic
equations we should present a connections between the momentum density
$\textbf{P}$ and the velocity field $\textbf{v}$. In the relativistic
hydrodynamics (RHD) we have two equations instead of the Euler
equation in the non-relativistic hydrodynamics, where the Euler equation
is both the momentum balance equation and the equation of evolution of
the particles current $\textbf{j}$ (which emerges in the continuity
equation and proportional to the velocity field
$\textbf{j}=n\textbf{v}$, where $n$ is the concentration of
particles). In the RHD's these equations are different. We can use
particles current evolution instead of evolution of momentum
density. So, we do not need to find connection between
$\textbf{P}$ and $\textbf{v}=\textbf{j}/n$. In the Newton's
mechanics and it's relativistic generalization the law of momentum
evolution is the basic dynamical law. It gives us microscopic
dynamical picture. When we are going to derive macroscopic
dynamical equations we should find one containing information
keeping in the Newton's second law, and the particles current
evolution equation is one of them, as well as the momentum balance
equation, these two equations match in the non-relativistic case.

When one has relativistic momentum balance equation it is a hard
job to find a connection between momentum density
$\textbf{P}(\textbf{r},t)$ and particles current
$\textbf{j}(\textbf{r},t)=n(\textbf{r},t)\textbf{v}(\textbf{r},t)$
to obtain the close set of the hydrodynamics equations for the system of
particles with the temperature. Where is the connection between
the momentum $\textbf{p}_{i}(t)$ and the velocity $\textbf{v}_{i}(t)$ of
the one particle
\begin{equation}\label{RCHD mom and vel of part
connection}\textbf{p}_{i}(t)=\frac{m_{i}\textbf{v}_{i}(t)}{\sqrt{1-\frac{v_{i}^{2}(t)}{c^{2}}}},\end{equation}
and
\begin{equation}\label{RCHD vel via momentum}\textbf{v}_{i}(t)=c\frac{\textbf{p}_{i}(t)}{p_{0i}(t)},\end{equation}
where $c$ is the speed of light, and $p_{i}^{0}$ is proportional
to the energy of the particle
\begin{equation}\label{RCHD energy via mom}p_{i}^{0}(t)c=E_{i}(t)=\sqrt{p_{i}^{2}(t)c^{2}+m_{i}^{2}c^{4}}.\end{equation}

If we consider system of particles, it is no simple matter to
suggest a connection between
$\textbf{P}=\sum_{i}m\textbf{v}_{i}(t)/\sqrt{1-v_{i}^{2}(t)/c^{2}}$
and $\textbf{v}=\sum_{i}\textbf{v}_{i}(t)$.

When we consider the relativistic dynamics the Newton's equation still
can be used, but we must consider the relativistic connection
(\ref{RCHD mom and vel of part connection}) between
$\textbf{p}_{i}(t)$ and $\textbf{v}_{i}(t)$, so
\begin{equation}\label{RCHD}\frac{d}{d t}\textbf{p}_{i}(t)=e\textbf{E}(\textbf{r}_{i}(t),t)+\frac{e}{c}\textbf{v}_{i}(t)\times \textbf{B}(\textbf{r}_{i}(t),t),\end{equation}
where $\textbf{E}(\textbf{r}_{i}(t),t)$ and
$\textbf{B}(\textbf{r}_{i}(t),t)$ are the electric and magnetic
fields acting on the i-th particle.
Otherwise, for particle's acceleration we have
$$\frac{d}{d t}\textbf{v}_{i}(t)=\frac{e}{m}\sqrt{1-\frac{v_{i}^{2}(t)}{c^{2}}}\times$$
$$\times\biggl(\textbf{E}(\textbf{r}_{i}(t),t)+\frac{1}{c}[\textbf{v}_{i}(t),\textbf{B}(\textbf{r}_{i}(t),t)]$$
\begin{equation}\label{RCHD}-\frac{1}{c^{2}}\textbf{v}_{i}(t)(\textbf{v}_{i}(t)\textbf{E}(\textbf{r}_{i}(t),t))\biggr).\end{equation}

In this paper we pay attention to the classic mechanics and
hydrodynamics, but where are also papers dedicated to the quantum
hydrodynamics ~\cite{Andreev PRB11}, ~\cite{Andreev PRA08},
~\cite{Shukla RMP 11}, and the relativistic quantum hydrodynamics
~\cite{Asenjo PP 11}, ~\cite{Haas PRE 12}.

This paper is organized as follows. In Sec. II we present a brief
review of the relativistic hydrodynamics for particles system with
the small thermal velocity spread. In Sec. III we consider
existing in the literature methods of obtaining of the RHD
equations for the system of relativistic particles with the
temperature. In Sec. IV we derive the RHD equations from the
picture of microscopic motion of particles described by the
Newton's law. In Sec. V we consider temporal evolution of new
function appeared at derivation described in the Sec. IV. In Sec.
VI we derive the equation of evolution for the energy-momentum
density to show that our treatment give the well-known equation.
In Sec. VII we discuss ways to obtain a close set of equations. In
Sec. VIII we consider dispersion of waves in the relativistic
plasma. In Sec. IX we present the brief summary of our results.

\section{\label{sec:level1} II. Relativistic hydrodynamics of particles system at small temperature}

As a simple example of relativistic many particle system we
consider a relativistic electron beam. If it's beam is
monoenergetic or it has small thermal spread of particle
velocities when all particles have (near) equal velocities
$\textbf{u}$, so momentum of the system
$\textbf{P}=\sum_{i=1}^{N}m\textbf{u}/\sqrt{1-u^{2}/c^{2}}=Nm\textbf{u}/\sqrt{1-u^{2}/c^{2}}$
is proportional to the momentum of single particle, and it
connects with the particle velocity by well-known relativistic
formula. Here and below, we use $\textbf{u}$ as velocity field for
approximately monoenergetic system of particles. Here and below,
for simplicity, we consider one species.

Following by Ref.s ~\cite{Hazeltine APJ 02}, ~\cite{Bret APJ 09},
~\cite{Bret PP 06}, ~\cite{Bret PP 08}, we can write an example of
using in literature set of the RHE
\begin{equation}\label{RCHD continuity n,v}\partial_{t}n+\nabla(n \textbf{v})=0,\end{equation}
and
\begin{equation}\label{RCHD mom evol simple}\partial_{t}\textbf{p}+(\textbf{v}\nabla)\textbf{p}=e\textbf{E}+\frac{e}{c}\textbf{v}\times\textbf{B},\end{equation}
where momentum of medium is presented in the form
$\textbf{p}=\gamma m\textbf{v}$, where
$\gamma=(1-v^{2}/c^{2})^{-1/2}$, analogously to single particle,
to close the set of equations. In equations (\ref{RCHD continuity
n,v}) and (\ref{RCHD mom evol simple}) $n$ is the concentration of
particles, $\textbf{v}$ is the velocity field,
$\textbf{p}=\textbf{P}/n$, where $\textbf{P}$ is the density of
momentum, $\textbf{E}$ and $\textbf{B}$ are the electric and
magnetic fields.

\section{\label{sec:level1} III. Short review of basic points of relativistic hydrodynamics}

In well-known course of theoretical physics written by L. D.
Landau and E. M. Lifshitz ~\cite{Landau v6}, ~\cite{Landau v2} we can find a
macroscopic derivation of the relativistic hydrodynamics for the
system of neutral particles. This derivation based on treatment of
the macroscopic bit of medium. They start their consideration from
definition of the stress-energy tensor $\hat{T}$ in the rest
frame. In this case they get
\begin{equation}\label{RCHD mom-en tensor in the rest frame}\hat{T}=
\left(\begin{array}{cccc}e&
0&
0&
0\\
0&
p&
0&
0\\
0 &
0 &
p&
0\\
0&
0&
0&
p\\
\end{array}\right),\end{equation}
where $e$ is the inner energy density, $p$ is the pressure, these
quantities are defined in the rest frame, i.e. the frame (may be
local) where is no macroscopic motion of the medium. Having
(\ref{RCHD mom-en tensor in the rest frame}) they use the Lorentz
transformation to get in an arbitrary inertial frame. In the result they find
\begin{equation}\label{RCHD mom-en tensor gen LL}T^{ab}=hu^{a}u^{b}/c^{2}+p g^{ab},\end{equation}
where $g^{ab}$ is  the metric sign convention such that $g^{ab} =
diag (1,-1,-1,-1)$, and $h=e+p$ is the enthalpy density, and Latin indexes $a$, $b=0$, 1, 2, 3.

The stress-energy tensor contains information about density of
dynamical quantities describing the system, consequently, its can
be used to write equation evolution of the medium which is an
analog of the non-relativistic Euler equation, and we also should
write the continuity equation which we present here in the terms
of the four-dimensional variables
\begin{equation}\label{RCHD part number cons law} \frac{\partial j^{a}}{\partial x^{a}}=0\end{equation}
is the number of particles conservation law, where $j^{a}$ is the
four-current.
\begin{equation}\label{RCHD en-mom cons law} \frac{\partial T^{ab}}{\partial x^{b}}=0\end{equation}
is the stress-energy conservation law. In formulas (\ref{RCHD part
number cons law}) and (\ref{RCHD en-mom cons law}) $x^{a}=(ct, x,
y, z)$, where $t, x, y, z$ are the independent variables, whereas
in the microscopic mechanics we have (x(t), y(t), z(t), t).

To write (\ref{RCHD mom-en tensor in the rest frame}) we should
consider a local frame where this piece of medium does not move.
Generally speaking this piece can be involved in complex motion,
and moving with acceleration relatively to some global inertial
frame. Moreover, if the piece is not moving macroscopically in chosen inertial
frame, its particles in their thermal motion can move with
relativistic velocities. Thus, we can not use non-relativistic
formulas in this frame too. Presented in the next section
description is free of these restrictions.


More general approximation in compare with the described above was developed by S. M. Mahajan in 2003
~\cite{Mahajan PRL 03}, and effectively used in many papers, for
example we present several of them ~\cite{Mahajan PRL 08},
~\cite{Asenjo PP 09}, ~\cite{Asenjo PRE 12}. This approximation
allows to get "minimal coupling" for the system of relativistic
charged particles. It includes thermal distribution of particles
and founds on the Lorentz invariance of equation of motion. When
we consider system of charged particles, in right-hand side of the
equation (\ref{RCHD en-mom cons law}) the electromagnetic field
emerges. Therefore, instead of (\ref{RCHD en-mom cons law}) we
have
\begin{equation}\label{RCHD mom-en tensor gen LL and EM field}\frac{\partial T^{ab}}{\partial x^{b}}=F^{ab}j_{b},\end{equation}
where $F^{ab}=\partial^{a}A^{b}-\partial^{b}A^{a}$ is the tensor
of the electromagnetic field and $A^{a}$ is the four-potential of the
electromagnetic field. Introducing $\textbf{p}-(e/c)\textbf{A}$
instead of $\textbf{p}$ we can rewrite (\ref{RCHD mom-en tensor
gen LL and EM field}) as
\begin{equation}\label{RCHD mom-en tensor gen LL and EM field short}\frac{\partial \textbf{T}^{ab}}{\partial x^{b}}=0,\end{equation}
where $\textbf{T}^{ab}$ is the sum of the stress-energy tensor
$T^{ab}$ and the electromagnetic stress-energy tensor
$T^{ab}_{EM}$, where
$$T^{ab}_{EM}=\frac{1}{4\pi}\biggl(F^{ac}g_{cd}F^{bd}-\frac{1}{4}g^{ab}F^{cd}F_{cd}\biggr)$$

In some cases, see for example sect. II of this paper, we can
present $T^{ab}$ via four-velocity $U^{a}=(\gamma, \gamma
\textbf{u}/c)$ as
\begin{equation}\label{RCHD}T^{ab}=\partial^{a}U^{b}-\partial^{b}U^{a}.\end{equation}
To account the statistical information of the particles system in
this formula in Ref. ~\cite{Mahajan PRL 03} a new tensor
$S^{ab}=\partial^{a}(fU^{b})-\partial^{b}(fU^{a})$ was introduced.
It contains parameter $f$, which is the  function of the temperature
$T$. Thus, in Ref. ~\cite{Mahajan PRL 03} a following equation was
suggested
\begin{equation}\label{RCHD MAH}U_{b}M^{ab}=0,\end{equation}
where $M^{ab}$ is the tensor that couples the electromagnetic and
the fluid fields, $M^{ab}=F^{ab}+(mc^{2}/e)S^{ab}$.

Following by the Ref.s ~\cite{Asenjo PP 09} we can represent the
spacelike components of equation (\ref{RCHD MAH})
\begin{equation}\label{RCHD}(\partial_{t}+\textbf{u}\nabla)(f\gamma \textbf{u})=\frac{e}{m}(\textbf{E}+\frac{\textbf{u}}{c}\times \textbf{B})-\frac{1}{mn\gamma}\nabla p.\end{equation}
We also represent corresponding continuity equation
\begin{equation}\label{RCHD cont with gamma}\partial_{t}(\gamma n)+\nabla(\gamma n \textbf{u})=0.\end{equation}
Usually equation (\ref{RCHD cont with gamma}) considered as the
continuity equation in the arbitrary frame. From this point of
view equation (\ref{RCHD continuity n,v}) is the continuity
equation in rest frame. However, below we show that at microscopic
derivation of the continuity equation in arbitrary inertial frame
we get (\ref{RCHD continuity n,v}).

Let us repeat the problem we discuss in the paper: we want to find
the form of the RHD equations for the system of particles with the
temperature, so, where there is distribution of particles by
velocities, and, in general, particles can move with relativistic
thermal velocities. One of the ways to solve with problem we just
describe following to the Ref. ~\cite{Mahajan PRL 03}, but where
was suggested the minimal coupling model. It includes the contribution of the
thermal motion in the RHD's, and it does not change  number of
equations (and variables) in the model. However, it is interesting
to see a way of widening of the model, a way of it's further
development beyond of the minimal coupling. Besides, one more crucial
moment, we do not want to make connection between the density of the
momentum $\textbf{P}(\textbf{r},t)$ and the velocity field
$\textbf{v}(\textbf{r},t)$ for the system of relativistic
particles with the temperature, but we want to be able to solve
described problem.

Another way to find equation evolution of the four-momentum is a use
of a kinetic equation.

In the physical kinetics the distribution function $f(\textbf{r},\textbf{p},t)$
is defined in the six dimensional space of possible value of the
coordinate $\textbf{r}$ and momentum $\textbf{p}$ of individual
particles. An individual particle's momentum and velocity connect
by formula (\ref{RCHD mom and vel of part connection}).
Introducing the distribution function as number of particles in the
vicinity of a point $\textbf{r}$ of the physical space and having
momentum in the vicinity of a point $\textbf{p}$ of the momentum space,
following to the Ref. ~\cite{Kuz'menkov 91}, we can write
$$f(\textbf{r},\textbf{p},t)=$$
$$\int_{\Delta_{r}}d\xi\int_{\Delta_{p}}d\eta \sum_{i=1}^{N}\delta(\textbf{r}+\xi-\textbf{r}_{i}(t))\cdot\delta(\textbf{p}+\eta-\textbf{p}_{i}(t)).$$
From this formula we see that $\textbf{p}$ connected with the momentum
of individual particles $\textbf{p}_{i}(t)$. Thus, in the kinetic
equation we can write $\textbf{p}=\gamma m\textbf{v}$. It is
well-known that the hydrodynamic functions can be found using the
distribution function, for example:
$$n(\textbf{r},t)=\int dp f(\textbf{r},\textbf{p},t),$$
$$\textbf{j}(\textbf{r},t)=\int dp \frac{\textbf{p}}{p_{0}} f(\textbf{r},\textbf{p},t).$$
In this case, one should use the kinetic equation to derive the
hydrodynamic equations (see for example Ref. ~\cite{Landau v10}).

\section{\label{sec:level1} IV. microscopic derivation of relativistic hydrodynamics equations for charged particles}

To start the derivation of the RHD equations we choose the
inertial frame. Next, we consider a sphere around each point of
space. Each moment of time we can calculate a number of particles
(or total mass of particles) in each sphere.

To find the concentration of particles in the vicinity of a point of the three dimensional physical space we should count the number of the particles and divide it on the volume of the vicinity
\begin{equation}\label{RCHD}\rho(\textbf{r},t)=\frac{1}{\Delta}\sum_{i=1}^{N(\textbf{r},t)}m_{i}. \end{equation}
Microscopic number of particles (total mass) in the vicinity of
the point $\textbf{r}$ changes during the time. This sum also
changes from one point of space to another. It is not suitable to
work with the sum which up limit of summation depend on
$\textbf{r}$ and $t$. Using the Dirac's delta function we can rewrite
$\rho(\textbf{r},t)$ in the following way
\begin{equation}\label{RCHD part num}\rho(\textbf{r},t)=\frac{1}{\Delta}\int d\xi\sum_{i=1}^{N}m_{i}\delta(\textbf{r}+\xi-\textbf{r}_{i}(t)).\end{equation}
If we consider system of particles with equal masses we can write
$$\rho(\textbf{r},t)=mn(\textbf{r},t),$$
where $n(\textbf{r},t)$ is the concentration of particles.

Differentiating the particles concentration with respect to time we obtain the continuity equation
\begin{equation}\label{RCHD cont}\partial_{t}n+\nabla \textbf{j}=0.\end{equation}
We do not chose any particular inertial frame. One might consider
the rest frame, but we start from the microscopic description, and the
notion "rest frame" does not clear from this point of view. Separate
particles have no information about: do they part of the
macroscopically motionless system or not? This question might be
answered on macroscopic scale only!

Current of particles appears in continuity equation, it's evident form is
$$\textbf{j}(\textbf{r},t)=\frac{1}{\Delta}\int d\xi\sum_{i=1}^{N}\textbf{v}_{i}\delta(\textbf{r}+\xi-\textbf{r}_{i}(t))$$
\begin{equation}\label{RCHD current}=\frac{1}{\Delta}\int d\xi\sum_{i=1}^{N}c\frac{\textbf{p}_{i}}{p_{0i}}\delta(\textbf{r}+\xi-\textbf{r}_{i}(t)).\end{equation}

For the following it is suitable to introduce the velocity field
\begin{equation}\textbf{v}(\textbf{r},t)=\frac{\textbf{j}(\textbf{r},t)}{n(\textbf{r},t)}.\end{equation}

Differentiation quantity (\ref{RCHD current}) with respect to time
we get simple, but very important equation: equation of particles
current evolution. One has very interesting form
\begin{equation}\label{RCHD bal of current}\partial_{t}j^{\alpha}+\partial^{\beta}\Pi^{\alpha\beta}=e\eta E^{\alpha}+\frac{e}{c}\varepsilon^{\alpha\beta\gamma} \eta^{\beta}B^{\gamma}-\frac{e}{c}\eta^{\alpha\beta}E^{\beta},\end{equation}
where Greek indexes $\alpha$, $\beta$ are used for 1, 2, 3; $\Pi^{\alpha\beta}$ is the current of $\textbf{j}$ or the
current of particles current \begin{equation}\label{RCHD curent of
curent}\Pi^{\alpha\beta}(\textbf{r},t)=\frac{1}{\Delta}\int
d\xi\sum_{i=1}^{N}v_{i}^{\alpha}v_{i}^{\beta}\delta(\textbf{r}+\xi-\textbf{r}_{i}(t)).\end{equation}
In the non-relativistic case $\Pi^{\alpha\beta}$ coincides with the
current of momentum.

In the equation of particles current evolution (\ref{RCHD bal of
current}) three new function appear. Evident form of them are
\begin{equation}\label{RCHD mu}\eta(\textbf{r},t)=\frac{1}{\Delta}\int d\xi\sum_{i=1}^{N}\frac{c}{p_{i}^{0}}\delta(\textbf{r}+\xi-\textbf{r}_{i}(t)),\end{equation}

\begin{equation}\label{RCHD mu--a}\eta^{\alpha}(\textbf{r},t)=\frac{1}{\Delta}\int d\xi\sum_{i=1}^{N}\frac{cv_{i}^{\alpha}}{p_{i}^{0}}\delta(\textbf{r}+\xi-\textbf{r}_{i}(t)),\end{equation}
and
\begin{equation}\label{RCHD mu--ab}\eta^{\alpha\beta}(\textbf{r},t)=\frac{1}{\Delta}\int d\xi\sum_{i=1}^{N}\frac{cv_{i}^{\alpha}v_{i}^{\beta}}{p_{i}^{0}}\delta(\textbf{r}+\xi-\textbf{r}_{i}(t)).\end{equation}

In the non-relativistic limit we find that $\eta\longrightarrow
n/m$, $\eta^{\alpha}\rightarrow j^{\alpha}/m$, and
$\eta^{\alpha\beta}\rightarrow \Pi^{\alpha\beta}/(mc)$. Comparing
the first and the third terms in the right-hand side of the equation
(\ref{RCHD bal of current}), including the evident form of the
concentration $n$ and $\Pi^{\alpha\beta}$, we can see that the
third term is  proportional to $v^{2}/c^{2}$, and exists in the
semi-relativistic approximation only. So, we should neglect it in the
non-relativistic limit, in compare with the first term. In the
result we have the usual non-relativistic Euler equation
$$m(\partial_{t}j^{\alpha}+\partial^{\beta}\Pi^{\alpha\beta})=en E^{\alpha}+\frac{e}{c}\varepsilon^{\alpha\beta\gamma} j^{\beta}B^{\gamma}.$$

We make our calculations in some inertial frame, but we do not
consider transition from one frame to another. For such transition
we can use the Lorentz transformation for the microscopic
quantities, and using  it we can find lows of transformation of
macroscopic quantities, but this topic lay of the paper.

Equation (\ref{RCHD bal of current}) and other dynamical equations
are written in the self-consistent field approximation, and we
also neglect by contribution of the electric dipole and other
moments of the medium. To do it we suppose that
$$E_{i}^{\alpha}(\textbf{r}_{i},t)=E_{i}^{\alpha}(\textbf{r}+\xi,t)\approx E_{i}^{\alpha}(\textbf{r},t),$$
and
$$B_{i}^{\alpha}(\textbf{r}_{i},t)=B_{i}^{\alpha}(\textbf{r}+\xi,t)\approx B_{i}^{\alpha}(\textbf{r},t).$$
A way of appearing of them is described in Ref.s ~\cite{Drofa TMP
96}, ~\cite{Andreev PIERS 2012} in the non-relativistic case. To
make the paper not to overload we left this topic for next one.

The electric $\textbf{E}$ and the magnetic $\textbf{B}$ fields arising
in equation (\ref{RCHD bal of current}) (and in other dynamical
equations presented in this paper) satisfy to the Maxwell's
equations
\begin{equation}\label{RCHD max 1}\nabla \textbf{B}=0,\end{equation}
\begin{equation}\label{RCHD max 2}\nabla \times \textbf{B}=\frac{1}{c}\partial_{t}\textbf{E}+\frac{4\pi e n \textbf{v}}{c},\end{equation}
\begin{equation}\label{RCHD max 3}\nabla \textbf{E}=4\pi en,\end{equation}
and
\begin{equation}\label{RCHD max 4}\nabla \times \textbf{E}=-\frac{1}{c}\partial_{t}\textbf{B}.\end{equation}

\section{\label{sec:level1} V. Evolution of the new functions}

Below we derive the equations of the momentum and the energy evolution,
but strictly speaking we do not need them to work with the RHD. In
fact we already have "equation of motion"-it is equation
(\ref{RCHD bal of current}).

Equation of the current evolution (\ref{RCHD bal of current}) introduce to us the three new functions ($\eta$, $\eta^{\alpha}$, and $\eta^{\alpha\beta}$) and we should find a way to close the set of the hydrodynamics equations to solve particular problems. One of possible ways to do it is to find equations of evolution of $\eta$, $\eta^{\alpha}$, and $\eta^{\alpha\beta}$. Certanly we can expect that in this case some new function will appear.

We have evident form of $\eta$ and $\eta^{\alpha}$, so, we can differentiate them with respect to time and obtain equations of these quantities evolution. In the result we have
\begin{equation}\label{RCHD bal of mu}\partial_{t}\eta+\partial_{\alpha}\eta^{\alpha}=-e\zeta^{\alpha}E^{\alpha},\end{equation}
and
\begin{equation}\label{RCHD bal of mu--a}\partial_{t}\eta^{\alpha}+\partial_{\beta}\eta^{\alpha\beta}=e\zeta E^{\alpha}+\frac{e}{c}\varepsilon^{\alpha\beta\gamma}\zeta^{\beta}B^{\gamma}-2\frac{e}{c}\zeta^{\alpha\beta}E^{\beta}.\end{equation}
As we expected several new functions appear. They evident form is
presented in this section below. But, from these equations
(\ref{RCHD bal of mu}) and (\ref{RCHD bal of mu--a}) we find
connection between $\eta$, $\eta^{\alpha}$, and
$\eta^{\alpha\beta}$. From (\ref{RCHD bal of mu}) we see that
$\eta^{\alpha}$ is the current of $\eta$, and, equation (\ref{RCHD
bal of mu--a}) shows us that $\eta^{\alpha\beta}$ is the current
of $\eta^{\alpha}$.

In general, the current of a quantity $f$, which we designate as
$f^{\alpha}$, might be presented in the form of
\begin{equation}\label{RCHD current general} f^{\alpha}=f\cdot v^{\alpha},\end{equation}
where $v^{\alpha}$ is the velocity field introduced above.
However, we neglect by the thermal motion in this formula, but it is
very useful approximation. If we include the thermal motion in the formula
(\ref{RCHD current general}) it assume the form $f^{\alpha}=f\cdot
v^{\alpha}+\emph{f}^{\alpha}$, where $\emph{f}^{\alpha}$ present
contribution of the thermal motion. For example, we consider the
non-relativistic current of the momentum $\Pi^{\alpha\beta}$, which
coincides with the current of particle current. In this case
$\Pi^{\alpha\beta}\simeq j^{\alpha}v^{\beta}$, and knowing that
$j^{\alpha}=n v^{\alpha}$, we have $\Pi^{\alpha\beta}\simeq n
v^{\alpha}v^{\beta}$. Including contribution of the thermal motion we
can write $\Pi^{\alpha\beta}\simeq n
v^{\alpha}v^{\beta}+p^{\alpha\beta}$, where $p^{\alpha\beta}$ is
the tensor of pressure. Usually one consider the scalar pressure $p$,
which connects with the tensor in following form
$p^{\alpha\beta}=p \delta^{\alpha\beta}$, where
$\delta^{\alpha\beta}$ is the Kronecker delta.

Thus, we can approximately write $\eta^{\alpha}=\eta v^{\alpha}$
and $\eta^{\alpha\beta}=\eta^{\alpha} v^{\beta}/c=\eta v^{\alpha}
v^{\beta}/c$. In the result, to understand meaning of $\eta$,
$\eta^{\alpha}$, and $\eta^{\alpha\beta}$ we need to understand
meaning of $\eta$ only. To do it we should consider its
semi-relativistic approximation.

Here, we present evident form of three functions which arise in
equations (\ref{RCHD bal of mu}) and (\ref{RCHD bal of mu--a})
\begin{equation}\label{RCHD chi}\zeta(\textbf{r},t)=\frac{1}{\Delta}\int d\xi\sum_{i=1}^{N}\Biggl(\frac{c}{p_{i}^{0}}\Biggr)^{2}\delta(\textbf{r}+\xi-\textbf{r}_{i}(t)),\end{equation}
\begin{equation}\label{RCHD chi--a}\zeta^{\alpha}(\textbf{r},t)=\frac{1}{\Delta}\int d\xi\sum_{i=1}^{N}v_{i}^{\alpha}\Biggl(\frac{c}{p_{i}^{0}}\Biggr)^{2}\delta(\textbf{r}+\xi-\textbf{r}_{i}(t)),\end{equation}
and
\begin{equation}\label{RCHD chi--ab}\zeta^{\alpha\beta}(\textbf{r},t)=\frac{1}{\Delta}\int d\xi\sum_{i=1}^{N}v_{i}^{\alpha}v_{i}^{\beta}\Biggl(\frac{c}{p_{i}^{0}}\Biggr)^{2}\delta(\textbf{r}+\xi-\textbf{r}_{i}(t)).\end{equation}

It can be shown that $\zeta$, $\zeta^{\alpha}$, and $\zeta^{\alpha\beta}$
connect with each other in the same way as $\eta$,
$\eta^{\alpha}$, and $\eta^{\alpha\beta}$. Thus we have
$\zeta^{\alpha}=\zeta v^{\alpha}$ and $\zeta^{\alpha\beta}=\zeta
v^{\alpha} v^{\beta}/c$.

\section{\label{sec:level1} VI. Stress-energy tensor}

Usually, at the RHD's description one consider tensor of
energy-momentum. It appears at consideration of evolution of the
energy-momentum four-vector density. To show our description
coincides with well-known we derive the equation of the
energy-momentum four-vector density evolution. The stress-energy
tensor appears with.

Let's introduce the density of the momentum
$$\textbf{P}(\textbf{r},t)=\frac{1}{\Delta}\int d\xi\sum_{i=1}^{N}\textbf{p}_{i}\delta(\textbf{r}+\xi-\textbf{r}_{i}(t))$$
\begin{equation}\label{RCHD momentum}=\frac{1}{\Delta}\int d\xi\sum_{i=1}^{N}\frac{m_{i}\textbf{v}_{i}}{\sqrt{1-\frac{v_{i}^{2}}{c^{2}}}}\delta(\textbf{r}+\xi-\textbf{r}_{i}(t)).\end{equation}
This quantity is a part of the stress-energy tensor $T^{ab}$.
Their connection is
$$\textbf{T}^{0\alpha}=c\textbf{P}.$$

Time part of the four-momentum vector is the energy. So, we
present the density of the energy
\begin{equation}\label{RCHD energy}T^{00}(\textbf{r},t)=\frac{1}{\Delta}\int d\xi\sum_{i=1}^{N}c\cdot p_{i}^{0}\delta(\textbf{r}+\xi-\textbf{r}_{i}(t)).\end{equation}

As in the cases described above we should differentiate the
definition of a quantity to find an equation of it's evolution. So,
we present the equation of the momentum density evolution
\begin{equation}\label{RCHD bal of mom}\frac{1}{c}\partial_{t}T^{0\alpha}+\partial^{\beta}T^{\alpha\beta}=enE^{\alpha}+\frac{e}{c}\varepsilon^{\alpha\beta\gamma}j^{\beta}B^{\gamma}.\end{equation}
In this equation only one "new" quantity appears, it is the tensor of momentum current
\begin{equation}\label{RCHD momentum current}T^{\alpha\beta}(\textbf{r},t)=\frac{1}{\Delta}\int d\xi\sum_{i=1}^{N}\frac{cp_{i}^{\alpha}p_{i}^{\beta}}{p_{i}^{0}}\delta(\textbf{r}+\xi-\textbf{r}_{i}(t)).\end{equation}

In the same way we find the energy density evolution equation
\begin{equation}\label{RCHD bal of energy}\frac{1}{c}\partial_{t}T^{00}+\partial^{\alpha}T^{\alpha 0}=\frac{e}{c}\textbf{j}\textbf{E},\end{equation}
where $T^{\alpha 0}$ is the current of energy, which simply
connects with the momentum density
$$T^{\alpha 0}=T^{0 \alpha}.$$

Analogously to the non-relativistic case we can write
$T^{\alpha\beta}=P^{\alpha}v^{\beta}+p^{\alpha\beta}$, where
$p^{\alpha\beta}\simeq p\delta^{\alpha\beta}$ is the pressure
tensor and $p$ is the isotropic pressure. Thus we can rewrite
equation (\ref{RCHD bal of mom}) in the following form
\begin{equation}\label{RCHD evol of mom with scal pressure}\partial_{t}P^{\alpha}+\partial^{\beta}(P^{\alpha}v^{\beta}+p\delta^{\alpha\beta})=enE^{\alpha}+\frac{e}{c}\varepsilon^{\alpha\beta\gamma}j^{\beta}B^{\gamma}.\end{equation}

Evolution of the energy density $T^{00}$ and the momentum density
$\textbf{P}$ depends on the concentration $n$, the particles current
$\textbf{j}$, the electric $\textbf{E}$ and the magnetic $\textbf{B}$
fields. However, evolution of $n$, $\textbf{j}$, $\textbf{E}$, and
$\textbf{B}$ do not depend on $\textbf{P}$ and $T^{00}$.
Therefore, to have the closed set of the RHD equations we do not need to
consider $\textbf{P}$ and $T^{00}$ evolution, and put our
attention for $n$, $\textbf{j}$ and $\eta$.

If we put $p=0$ and $\textbf{P}=n\textbf{p}$ in equation (\ref{RCHD evol of mom with scal pressure}) and include equation (\ref{RCHD continuity n,v}) or (\ref{RCHD cont}) (which is the same) we find that equation (\ref{RCHD evol of mom with scal pressure}) mach with the (\ref{RCHD mom evol simple}).

At studying of the relativistic plasma an approximation is used for
the pressure $p$. In order to close the set of the RHD equation
one use the equation of state for an ideal gas: $p=nk_{B}T$, where
$k_{B}$ is the Boltzmann constant and $T$ is the temperature. It
we consider evident form of $\Pi^{\alpha\beta}$ and
$T^{\alpha\beta}$ we will see that they have the same tensor
structure, i.e. they both depend on $v_{i}^{\alpha}v_{i}^{\beta}$.
Thus, we can use for $\pi^{\alpha\beta}$, where
$\Pi^{\alpha\beta}=nv^{\alpha}v^{\beta}+\pi^{\alpha\beta}$, the
same approximation as for $p^{\alpha\beta}$. Consequently, we can
put $\pi^{\alpha\beta}=\phi\cdot\delta^{\alpha\beta}$ and
$\phi=\phi(n,T)$.

\section{\label{sec:level1} VII. Closing of set of the hydrodynamics equations}

During the paper we attain that we can describe relativistic
plasma by means of $n$, $j^{\alpha}$, $\eta^{\alpha}$,
$E^{\alpha}$, and $B^{\alpha}$. In this case we include some of
effects caused by temperature, but some of them we lost. To study
relativistic plasma with the large temperatures we should consider
contribution of thermal motion at least in $\eta^{\alpha}$ and
$\eta^{\alpha\beta}$. However, we suppose not to consider it in
the paper. We already introduced new quantity $\eta$, and we
should understand it's meaning. For this purpose we consider
semi-relativistic limit for $\eta$, find out it's contribution in
dispersion of waves and try to approximately express it via $n$
and $\eta$. Even after getting approximate connection between $n$,
$\textbf{v}$, and $\eta$, we suppose to consider $\eta$ as an
independent variable, along with concentration $n$ and velocity
field $\textbf{v}$.

Thus, the semi-relativistic approximation of $\eta$ has form $\eta=n/m-\Pi^{\alpha\alpha}/2mc^{2}$, where $\Pi^{\alpha\alpha}$ is the trace of the tensor of the current of the particles current (\ref{RCHD curent of curent}), $\Pi^{\alpha\alpha}=nv^{2}+3\phi$, where $\phi$ is the current of the particles current on the thermal velocities. We have got it using the formula
$$\frac{1}{p_{0 i}}=\frac{1}{m_{i}c}\biggl(1-\frac{v^{2}}{2c^{2}}\biggr),$$
which is the semi-relativistic approximation for the inverse time component of the four-momentum.

\section{\label{sec:level1} VIII. Dispersion of longitudinal waves in relativistic plasma}

We suggest that in equilibrium state the relativistic electron
plasma (we suppose that ions motionless) is described dy following
parameters $n_{0}$, $\eta_{0}$, and $\textbf{v}_{0}=0$. To study dynamics of small perturbation we use equations (\ref{RCHD cont}), (\ref{RCHD bal of current}), and (\ref{RCHD max 3}). For the first step we suggest that $\phi$ depends on concentration $n$ only, but below we will account that $\phi$ depends on $n$ and $\eta$. We notice that in non-relativistic limit $\phi$ becomes pressure $p$, and $\eta$ becomes concentration $n$, and dependence on $n$ and $\eta$ reduces to dependence on $n$.
Considering evolution of small perturbations around the
equilibrium state we can find it's dispersion dependencies, which
has form of
\begin{equation}\label{RCHD disp rel w n}\omega^{2}=4\pi e^{2}\eta_{0}+\biggl(\frac{\partial\phi}{\partial n}\biggr)_{0}k^{2},\end{equation}
we can see that $\eta_{0}$ appears instead of the equilibrium concentration $n_{0}$, and $\phi$ emerges instead of the pressure $p$. In the semi-relativistic limit from (\ref{RCHD disp rel w n})  we get
$$\omega^{2}=\frac{4\pi e^{2}n_{0}}{m}-\frac{6\pi e^{2}\phi_{0}}{mc^{2}}+\biggl(\frac{\partial\phi}{\partial n}\biggr)_{0}k^{2},$$
where we use that $\Pi^{\alpha\alpha}_{0}\simeq 0+3\phi_{0}$, since $\textbf{v}_{0}=0$. Consequently, in the non-relativistic limit we have
\begin{equation}\label{RCHD disp non rel}\omega^{2}=\frac{4\pi e^{2}n_{0}}{m}+3\frac{k_{B}T}{m}k^{2},\end{equation}
 in the second term the equation of state of the ideal gas was
used at the adiabatic condition with the rate of adiabat equals 3.

Including that $\phi$ depends on both concentration $n$ and $\eta$ we still get the formula (\ref{RCHD disp rel w n}). However if we consider the wave of particle concentration in an electron beam we obtain dispersion equation
\begin{equation}\label{RCHD disp rel w n and eta}(\omega-kU)^{2}-\frac{1}{m}k^{2}\partial_{n}\phi+\frac{4\pi e^{2}\eta_{0}^{2}Uk}{mn_{0}(\omega-kU)}\partial_{\eta}\phi-4\pi e^{2}\eta_{0}=0,\end{equation}
which describes the longitudinal waves in the electron beam.
It is the equation of the third degree in contrast to (\ref{RCHD disp rel w n}) which is equation of the second degree.

Thus we can see that an account of $\phi(\eta)$ dependence could lead to some new effects. However, to get solution we should present  equation of state $\phi=\phi(n, \eta, T)$, which has not found yet. Presented equations and it's consequence give us a lot of new open questions, but they also give another view on the development of the RHD.

\section{\label{sec:level1} IX. Conclusion}

We have presented microscopic derivation of the relativistic
hydrodynamics equation. We have derive as well as well-known and
new equations. Among well-known equation we can mention continuity
equation, momentum balance equation and energy balance equation.
We have presented the particles current evolution equation, since the
particles current simply related with the velocity field
$\textbf{j}=n\textbf{v}$. At this equation derivation new
functions have appeared. To understand their meaning and to
consider their influence on particles dynamics we have derive
equations of theirs evolution. We have suggested the closed set of the
RHD equations and used its to consider dispersion of the collective
excitations in the relativistic plasma.



\begin{thebibliography}{99}

\bibitem{Hakim book 11} R. Hakim. "Introduction to
Relativistic Statistical Mechanics -- Classical and Quantum" (WS, 2011)

\bibitem{Kuz'menkov 91} L. S. Kuz'menkov, Theoretical and Mathematical
Physics \textbf{86}, 159 (1991).


\bibitem{Andreev PRB11} P. A. Andreev, L. S. Kuz'menkov, M. I.
Trukhanova, Phys. Rev. B \textbf{84}, 245401 (2008).

\bibitem{Andreev PRA08} P. A. Andreev, L. S. Kuz'menkov, Phys. Rev. A \textbf{78}, 053624 (2008).

\bibitem{Shukla RMP 11} P. K. Shukla, B. Eliasson,
Rev. Mod. Phys. \textbf{83},  885 (2011).

\bibitem{Asenjo PP 11} F. A. Asenjo, V. Munoz, J. A. Valdivia, and S. M.
Mahajan, Phys. Plasmas \textbf{18}, 012107 (2011).

\bibitem{Haas PRE 12} F. Haas, B. Eliasson, P. K. Shukla, Phys. Rev. E \textbf{85}, 056411 (2012).

\bibitem{Hazeltine APJ 02} R. D. Hazeltine and S. M. Mahajan, ApJ, \textbf{567}, 1262 (2002).

\bibitem{Bret APJ 09} A. Bret, ApJ, \textbf{699}, 990 (2009).

\bibitem{Bret PP 06} A. Bret, C. Deutsch, Physics of Plasmas, \textbf{13}, 042106 (2006).

\bibitem{Bret PP 08} A. Bret, M. E. Dieckmann, Physics of Plasmas, \textbf{15}, 062102 (2008).

\bibitem{Landau v6} L. D. Landau, E.M. Lifshitz
"Hydrodynamics" , Vol. 6 of Course of Theoretical Physics
(Pergamon, London, 1959).

\bibitem{Landau v2} L. D. Landau, E.M. Lifshitz
"Field theory" , Vol. 2 of Course of Theoretical Physics
(Pergamon, London, 19????).

\bibitem{Mahajan PRL 03} S. M. Mahajan, Phys. Rev. Lett. \textbf{90}, 035001 (2003).

\bibitem{Mahajan PRL 08} S. M. Mahajan, Phys. Rev. Lett. \textbf{100}, 075001 (2008).

\bibitem{Asenjo PP 09} F. A. Asenjo,  V. Munoz, J. A.Valdivia, T. Hada, Physics of
Plasmas, \textbf{16}, 122108 (2009).



\bibitem{Asenjo PRE 12} F. A. Asenjo, F. A. Borotto, A. C.-L. Chian, V. Munoz, J. A.Valdivia,
E. L. Rempel, Phys. Rev. E, \textbf{85}, 046406 (2012).



\bibitem{Landau v10} L. D. Landau, E.M. Lifshitz
"Physical kinetics" , Vol. 10 of Course of Theoretical Physics
(Pergamon, London, 1981).

\bibitem{Drofa TMP 96}  M. A. Drofa, L. S. Kuz'menkov,
Theoretical and Mathematical Physics \textbf{108}, 849 (1996).

\bibitem{Andreev PIERS 2012} L. S. Kuz'menkov  and P. A.  Andreev,
will be presented in PIERS Proceedings,  Augoust 19-23, Moscow,
Russia 2012.






\end{thebibliography}
\end{document}